# Ultra-sensitive magnetometry based on free precession of nuclear spins

C. Gemmel, W. Heil[a], K. Lenz, Ch. Ludwig, K. Thulley, Yu. Sobolev[b]
*Institut für Physik, 55099 Mainz, Germany*
M. Burghoff, S. Knappe-Grüneberg, W. Kilian, W. Müller, A. Schnabel, F. Seifert, L. Trahms
*Physikalisch-Technische Bundesanstalt, 10587 Berlin, Germany*
St. Baeßler[c]
*University of Virginia, Charlottesville, VA 22904, U.S.A*

**Abstract**
We discuss the design and performance of a very sensitive low-field magnetometer based on the detection of free spin precession of gaseous, nuclear polarized $^3$He or $^{129}$Xe samples with a SQUID as magnetic flux detector. The device will be employed to control fluctuating magnetic fields and gradients in a new experiment searching for a permanent electric dipole moment of the neutron as well as in a new type of $^3$He/$^{129}$Xe clock comparison experiment which should be sensitive to a sidereal variation of the relative spin precession frequency. Characteristic spin precession times $T_2^*$ of up to 60h could be measured. In combination with a signal-to-noise ratio of > 5000:1, this leads to a sensitivity level of $\delta B \approx 1 fT$ after an integration time of 220s and to $\delta B \approx 10^{-4} fT$ after one day. Even in that sensitivity range, the magnetometer performance is statistically limited, and noise sources inherent to the magnetometer are not limiting. The reason is that free precessing $^3$He ($^{129}$Xe) nuclear spins are almost completely decoupled from the environment. That makes this type of magnetometer in particular attractive for precision field measurements where a long-term stability is required.

# I. Introduction

Magnetometers are intended for precise measurement and control of magnetic fields and magnetic field fluctuations on a very broad dynamic range extended from strong magnetic fields (a few Tesla) down to very small magnetic fields (fT). For the past 30 years, superconducting quantum interference devices (SQUIDs) operating at 4K have been unchallenged as ultrahigh-sensitivity magnetic field detectors, with a sensitivity reaching down to $1\, fT/\sqrt{Hz}$ [1]. In recent years, however, significant technical advances have enabled atomic magnetometers to achieve sensitivities rivalling [2-4] and even surpassing [5] that of most SQUID-based magnetometers. The instruments developed and commercially available have already a high performance. Nonetheless, the research on new magnetometers with improved performance and for a broader range of applications is still a very active area.

---
[a] Corresponding author: wheil@uni-mainz.de
[b] on leave from PNPI Gatchina, Russia
[c] former member of the Institute of Physics, Mainz

The magnetometer described here is based on the detection of free spin-precession of gaseous, nuclear spin-polarized $^3$He or $^{129}$Xe samples with a SQUID as magnetic flux detector. Such type of magnetometer has been achieved in the past by Cohen-Tannoudji et al. [6]. They already performed ultra-sensitive ($\sim 100\, fT/\sqrt{Hz}$) magnetometry with a $^{87}$Rb-magnetometer using the ground-state Hanle effect. The modulation of the magnetic field due to the free precession of the $^3$He nuclear spins could be recorded for several hours with a measured transverse nuclear relaxation time $T_2^*$ of $T_2^* = 140$ min.

The overall sensitivity of such a $^3$He-SQUID or $^3$He-Rb magnetometer can be estimated using the statistical signal processing theory [7]: For a sinusoidal magnetometer signal, the frequency $f$ is to be determined from the recorded data points. Those can be written as:

$$S[n] = A \cdot \cos(2\pi \cdot (f/r_s) \cdot n + \Phi) + w[n] \qquad n = 0,1,2,3,...,N-1 \tag{1}$$

where $\Phi$ is the initial phase, and $w[n]$ is the white Gaussian noise.

For detection times $T \leq T_2^*$, where the exponential damping of the recorded free precession signal ($A$) affects the sensitivity of the magnetometer not too much, we can introduce an average value ($\overline{SNR} = \overline{A}/N_\alpha$) of the measured signal-to-noise ratio. The noise is defined as the square root of the integrated power spectral density $\rho_\alpha^2$ of the corresponding signal fluctuations

$$N_\alpha = \left( \int_0^{f_{BW}} \rho_\alpha^2 \cdot df \right)^{1/2} \tag{2}$$

where $f_{BW} = r_s/2$ is the sampling rate ($r_s$) limited bandwidth, i.e., the Nyquist frequency. If the noise is white, the noise level is given by $N_\alpha = \rho_\alpha \sqrt{f_{BW}}$.

According to ref.[7], the Cramer-Rao Lower Bound (CRLB) sets the lower limit on the variance $\sigma_f^2$ of any frequency estimator:

$$\sigma_f^2 \geq \frac{12}{(2\pi)^2 \cdot (\overline{A}/N_\alpha)^2 \cdot f_{BW} \cdot T^3} \tag{3}$$

From continued recording of the precessing sample magnetization, the measurement sensitivity on $f$ or, by using

$$f = \gamma/(2\pi) \cdot B_0, \tag{4}$$

the sensitivity $\delta B$ on the respective magnetic field $B_0$ seen by the sample spins increases with the observation time, $T$, according to

$$\delta B \geq \frac{\sqrt{12}}{(\bar{A}/\rho_\alpha) \cdot T^{3/2} \cdot \gamma} ,  \quad (5)$$

where $\gamma$ is the gyromagnetic ratio.

In the appendix, an improvement to the sensitivity estimate is given that takes the exponential damping of the precession signal into account.

Due to their three orders of magnitude higher gyromagnetic ratio $\gamma$, equation (5) suggests to use magnetometers based on the spin precession of electrons rather than on the spin precession of nuclei. Indeed, the atomic magnetometer with the best short-term sensitivity is the spin-exchange-relaxation-free (SERF) magnetometer whose sensitivity exceeds $fT/\sqrt{Hz}$ in practice [8]. For long-term magnetic field measurements, however, one can make use of the $T^{-3/2}$ decrease of $\delta B$, provided the coherent spin-precession survives long enough. Usually, the relaxation time of electron spins is short, while nuclei, such as $^3$He, display a much longer spin-relaxation time. This may make them competitive or even superior to electron-spin magnetometers. In searches for non-magnetic spin interactions the smallness of the nuclear moment is actually an advantage, as it reduces the sensitivity to spurious magnetic effects.

The Allan Standard Deviation [9] is the most convenient measure to study the temporal characteristics of magnetometers. However, since external field fluctuations are the dominant sources of magnetic noise, in that sensitivity range, limitations, and thus deviations from the CRLB power law, due to noise sources inherent to the magnetometer proper are difficult to determine. The way out is to use co-located magnetometers, e.g., a $^3$He/$^{129}$Xe clock, where one compares the transition frequencies of two co-located magnetometers. In this case, the Zeeman-term drops out to first order and the Allan Standard Deviation is then used to study the characteristics of the frequency- or phase noise error (Section 4.2).

While an atomic magnetometer, e.g., Cs-magnetometer, uses the transition frequency between two energy levels in an atom, the environment, which the quantum absorber (atom) is exposed to, can perturb the energies of these two energy states. For example, it is well known that a near-resonant light field shifts the Zeeman levels in the same way as a static magnetic field oriented along the light beam (excess noise $\delta B_{LS}$). The AC-Stark shift, and hence $\delta B_{LS}$ is proportional to the light intensity and has a dispersive dependence on the detuning of the laser frequency from the optical absorption line. Thus light power fluctuations may limit the ultimate sensitivity of such magnetometers as discussed, e.g., in ref. [4].
In case of free precessing nuclear spins, the environment is almost completely decoupled (up to the small effect of chemical shifts that does not led to additional excess noise ) that makes this type of magnetometer in particular attractive for precision field measurements where a long-term stability is required.

Since the pioneering works of C.Cohen-Tannoudji et al., a lot of improvements have been made on the parameters which determine the measurement sensitivity of a free precession $^3$He ($^{129}$Xe) magnetometer: Firstly, thanks to the use of lasers instead of discharge lamps for the optical pumping of $^3$He ($^{129}$Xe) [10,11], the sample polarization could be increased from 5% (typically) to up to 90% [12]. Secondly, the use of low-$T_c$ DC-SQUID magnetometers as magnetic flux detectors with a white magnetic noise level of $\rho_{SQUID} \approx 2\, fT/\sqrt{Hz}$ [13,14,15].

Thirdly, as will be shown in more detail in Section 3, the transverse spin relaxation time $T_2^*$ could be increased by more than an order of magnitude using low-relaxation glass containers for the polarized $^3$He ($^{129}$Xe) samples, immersed in homogeneous magnetic guiding fields of ≤ 1μT inside strongly magnetically shielded rooms .

The paper is organized as follows: In Section 2 we discuss ways to obtain long nuclear-spin phase coherence times, i.e., long $T_2^*$-times, along with the basic layout of the experimental setup. The study of free precession of $^3$He nuclear spins in a spherical sample cell is presented in Section 3. Two actual applications of this type of $^3$He- or $^3$He/$^{129}$Xe-SQUID magnetometer will be presented in Section 4, where long-term stability in the sub fT sensitivity range is of decisive importance:  (i) The main factors limiting the present-day accuracy in experiments searching for the electric dipole moment (EDM) of the neutron are the precise measurement and control of magnetic fields and gradients. In Section 4.1 a large area $^3$He magnetometer is presented which meets the requirements of upcoming neutron EDM experiments. (ii) In order to be insensitive to magnetic effects (Zeeman-term), co-located atomic clocks are used to search for new physics, which may violate Lorentz symmetry and/or CPT invariance. In Section 4.2 we discuss a new type of $^3$He/$^{129}$Xe clock comparison experiment, which should be sensitive to a sidereal variation of the relative spin precession frequency.  Conclusion and outlook (Section 5) is followed by an appendix (A) with the derivation of the CRLB frequency estimate for an exponentially damped sinusoidal signal.

## 2. Concept of long nuclear-spin phase coherence times and basic layout of experimental setup

### Methodical Background

The presence of a magnetic field gradient in a sample cell containing spin-polarized $^3$He ($^{129}$Xe) gas will cause an increased transverse relaxation rate. The origin of this relaxation mechanism is the loss of phase coherence of the atoms due to the fluctuating magnetic field seen by the atoms as they diffuse throughout the cell. Based on the Redfield theory of relaxation [16] due to randomly fluctuating magnetic fields, analytical expressions can be derived for the transverse relaxation rate for spherical and cylindrical sample cells, as reported by G.D.Cates et al. [17] and D.D.McGregor [18], respectively. Taking into account the relaxation rate at the walls, $1/T_{1,wall}$, and other spin-relaxation modes subsumed under the longitudinal relaxation time $T_1$, the general expression for the transverse relaxation rate $1/T_2^*$ for a spherical sample cell of radius $R$ is

$$\frac{1}{T_2^*} = \frac{1}{T_1} + \frac{1}{T_{2,field}} = \frac{1}{T_1} + \frac{8R^4\gamma^2\left|\vec{\nabla}B_{1,x}\right|^2}{175\,D} + D\frac{\left|\vec{\nabla}B_{1,y}\right|^2 + \left|\vec{\nabla}B_{1,z}\right|^2}{B_0^2} \cdot \sum_n \frac{1}{\left|x_{1n}^2 - 2\right| \cdot \left(1 + x_{1n}^4\left(\gamma B_0 R^2/D\right)^{-2}\right)}$$

(6)

with the magnetic holding field pointing into the x-direction.

$D$ is the diffusion coefficient of the gas ($D_{He}$=1880 cm$^2$/s and $D_{Xe}$= 58 cm$^2$/s at 1 mbar and T=300K, [19]), and $x_{1n}$ (n=1,2,3,…) are the zeros of the derivative $(d/dx)j_1(x) = 0$ of the spherical Bessel function $j_1(x)$. The deviation $B_1(r)$ of the local field from the average homogeneous field $B_o$ was approximated by the uniform gradient field $B_1(r)=r\cdot\nabla B_1$, with $\nabla B_1$ being a traceless, symmetric second-rank tensor. The pressure dependence of the transverse relaxation rate ($1/T_2^*$) of $^{129}$Xe/N$_2$ gas mixtures was already studied quantitatively by some of the authors [14] at room temperature in ultra-low magnetic fields ($B_o = 4.5..15$ nT) inside a magnetically shielded room (BMSR-1 [20]). The results confirmed the predictions of the existing theory of spin-relaxation in the motional narrowing regime [17], i.e., $\tau_{diff}/\tau_L \ll 1$ (diffusion time $\tau_{diff}$ required for the spins to diffuse across the cell much smaller than the characteristic spin precession time $\tau_L$). At BMSR-1 as well as in the new magnetically shielded room (BMSR-2, [21]) at the Physikalisch Technische Bundesanstalt Berlin (PTB), the residual magnetic field gradients are of order pT/cm. A sophisticated demagnetization procedure in BMSR-2 preserves a low residual magnetic field of $|B_{res}| < 2 nT$ in a measurement volume of 1 m$^3$[22]. Equation (6) above suggests to measure at low pressures, the regime of motional narrowing, and at low magnetic fields in order to minimize the transverse relaxation rate, since the field gradient induced rate then gets proportional to the square of absolute field gradients and to $1/D \sim p$, i.e., proportional to the gas pressure ($p$):

$$\frac{1}{T_{2,field}} \approx \frac{4R^4\gamma^2}{175 D}\left(\left|\vec{\nabla}B_{1,y}\right|^2 + \left|\vec{\nabla}B_{1,z}\right|^2 + 2\left|\vec{\nabla}B_{1,x}\right|^2\right) \tag{7}$$

However, lowering both gas pressure and magnetic field, reduces the *SNR* and with it the measurement sensitivity according to equation (5). For the signal intensity we have $A \sim p$, and for spin precession frequencies of $\omega/2\pi = (\gamma/2\pi)\cdot B_0 < 10\ Hz$ we approach the spectral region of elevated signal noise (mechanical vibration, *1/f*-noise,…) as it is shown in Fig.1 and described in detail in ref.[23].

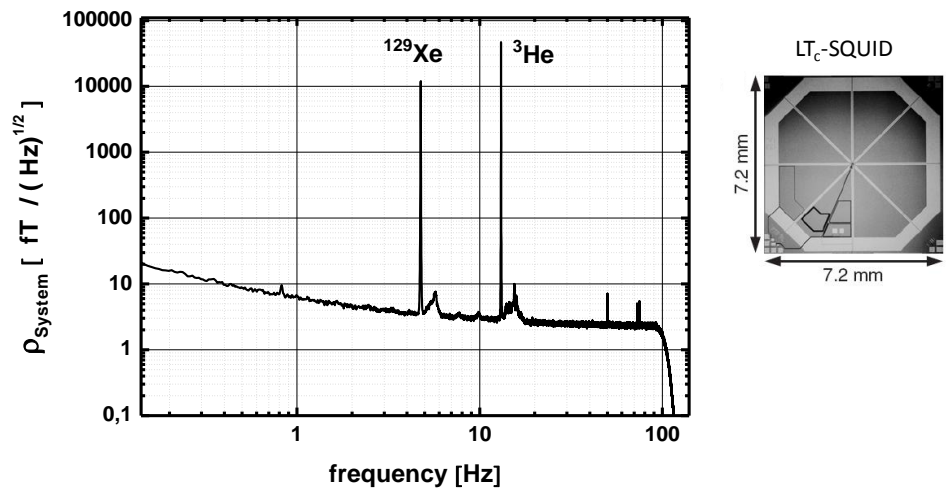

Fig.1: Magnetic flux density spectrum of a SQUID measuring the precession of $^3$He and $^{129}$Xe. The prominent features at about 4.7 Hz and 13 Hz correspond to the Larmor oscillation of the co-located $^{129}$Xe and $^3$He spins in one sample cell at a field of 400 nT. For frequencies $f > 10$ Hz, we find a white system noise of $\rho_{system} \approx 2.3 \, fT/\sqrt{Hz}$. The cut-off frequency is at 125 Hz (sampling rate: $r_s$ = 250 Hz). 'Bumps' and lines in the spectra at low frequencies are caused by mechanical vibrations or power line interference. At the right side a photograph of the low-$T_c$ multiloop SQUID magnetometer operated inside the PTB's magnetically shielded room is shown.

From these lower bounds, it can be inferred that optimum conditions are met at magnetic fields around 1 μT ($f_{He,Xe} \approx 10 \, Hz$) and at gas pressures around 1 mbar. The pressure range of 1 mbar is the required gas pressure for metastable optical pumping (MEOP) of $^3$He, as well, that facilitates the use of $^3$He based spin precession magnetometers with SQUID readout. In MEOP [24], the $^3$He atoms are excited into the metastable level $2^3S_1$ (~1ppm) by a weak, high frequency discharge. In this level, an electron polarization is obtained by absorption of a circularly polarized light beam ($\lambda$=1083 nm) whose frequency is tuned to excite, e.g., the optical transition from $2^3S_1$ ($F$=3/2) in $2^3P_0$ ($F$=1/2). Because of hyperfine coupling, the electronic orientation becomes a nuclear orientation. Finally, the metastability exchange collisions with ground state atoms ($1^1S_0$) transfer nuclear orientation to the ground state. The measurement of the static magnetic field can be made by recording the precessing magnetization of the nuclear spins. Let us consider a spherical cell with low-pressure $^3$He gas, which is homogeneously polarized. The magnetic field being produced outside the cell is the same as the field created by a dipole located at its centre and characterized by a magnetic moment $M_o$ given by

$$M_o = \mu_{He} \cdot P \cdot N \quad (8)$$

where $\mu_{He}$ is the $^3$He nuclear magnetic moment ($\mu_{He}$ = 1.08·10$^{-26}$ JT$^{-1}$), $P$ is the polarization, and $N$ is the number of atoms in the ground state. Along the dipole axis, at a distance $d$ from the centre, the field $\Delta B$ is then given by

$$\Delta B = \frac{\mu_o}{4\pi} \cdot \frac{2M_o}{d^3} \quad (9)$$

This results in a magnetic field at the surface of the sample cell of $\Delta B$[pT] ≈ 220·$p$·$P$ with $p$ in units of mbar (at room temperature). Taking $\rho_{SQUID} \approx 3 \, fT/\sqrt{Hz}$ for the SQUID noise (white noise), a first rough estimation shows that a *SNR* of *SNR*=$A/N_\alpha$ > 3000:1 in a bandwidth of 1 Hz can be easily reached for SQUIDs positioned close to the sample cell.

**Experimental setup**

The instrumental setup is sketched in Fig.2. The novel SQUID vector magnetometer system has been specially designed for biomagnetic applications inside the strongly magnetically shielded room BMSR-2 at the PTB [25,26]. It is housed in a Dewar with a flat bottom and an inner diameter of Ø = 250 mm. The SQUIDs are arranged so that in addition to the usually measured z-component of the field, the horizontal magnetic fields can be measured, too. A total of 304 DC-SQUID magnetometers are divided up into 19 identical modules. The 16 low-$T_c$ SQUIDs of each module are located in such a way that an estimation of the magnetic field in all three dimensions is possible inside the module. The 57 SQUIDs of the lowest z-plane

(1.5 cm above the Dewar bottom) of all modules form a hexagonal grid. In our measurements, we put the sample cells directly below the Dewar at one of the central modules and we refer to data recorded by the SQUIDs of its lowest plane in the vertical (-z)-direction.

Inside the µ-metal shielded room, a homogeneous magnetic field of about 400 nT was provided by two quadratic coil pairs ($B_x$-coil and $B_y$-coil) which were arranged perpendicular to each other (see Fig.2). The use of two coil pairs was chosen in order to manipulate the sample spins. A slow rotation of the magnetic field from $B_x$ to $B_y$ or vice versa causes an adiabatic rotation of the spins for $\Omega_{rot}/\omega_L \ll 1$ while for $\Omega_{rot}/\omega_L \gg 1$ the non-adiabatic condition is met. The latter measure was used to realize a "π/2-pulse". This way, nuclear spin-precession in the xz-plane or, alternatively, in the yz-plane could be monitored. According to equation (4), the corresponding Larmor precession frequencies in a field of $B_0 \approx 400$ nT are $\nu_{L,He} \approx 13$ Hz and $\nu_{L,Xe} \approx 4.7$ Hz, respectively. The $^3$He sample cells could be polarized either directly by MEOP inside the shield or glass cells were filled with polarized gases ($^3$He, $^{129}$Xe[1], or gas mixtures) from low-relaxation storage vessels outside the shielding in the so-called sample cell preparation area (see Fig.2).

In order to provide a well defined magnetic guiding field $B_G$ for the nuclear spins during transport into the magnetically shielded room, a transport coil of length $L$ consisting of an outer (o) and inner (i) solenoid with magnetic moments $M_O = -M_I = n_o \cdot I_o \cdot A_o$ was used with a resulting field of $B_G = B_i + B_0 = \mu_0 \cdot (n_i/2) \cdot I_i$ taking $n_o = n_i/2$, $I_o = -I_i$, and $A_o = 2A_i$. Since the axial stray fields of this double-solenoid system drop $\sim 1/z^5$, the fringe field of the transport coil reaches the 400nT level already after a distance of z ∼ 30 cm from the solenoid ($B_G \approx 0.3$ mT). This guaranteed that the inner µ-metal walls were not magnetized and that the field-gradients stayed almost constant over the individual measurement cycles. The magnetic field gradients with and without magnetic holding field in BMSR-2 were measured with the SQUID vector magnetometer system itself and are listed in Table 1. To evaluate the different tensor components of the gradient field, the SQUID magnetometer system and thus the Dewar as a whole had to be moved to preset positions forming a grid in 3D-space around the sample position. The main uncertainty in the determination of the components of the magnetic field gradient is the incorrect alignment of the Dewar and the fact that in presence of a magnetic guiding field ($B_x \neq 0$) this misalignment had a strong influence mainly on the extraction of the transverse components of the magnetic field gradient. In our analysis (see Section 3) we took the average value of the measured extreme values (see Table 1) and assigned their deviation as error bar (1σ).

---

[1] $^{129}$Xe was polarized by means of spin-exchange optical pumping [27]

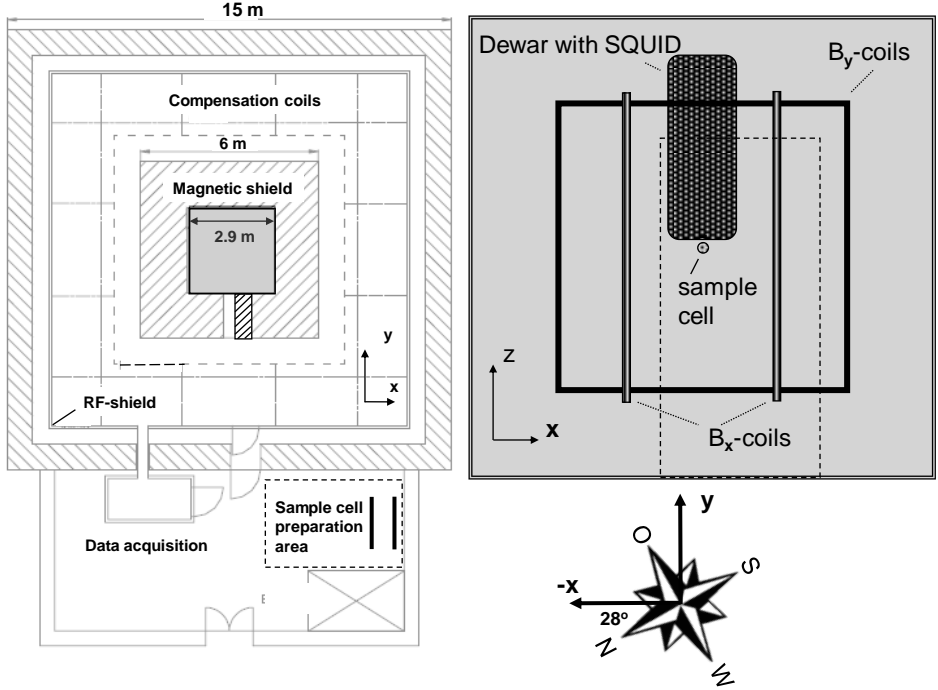

Fig.2: (Left) Horizontal cut view through building, shielded room and annex with data acquisition chamber and sample cell preparation area. The passive shielding factor of the BMSR-2 exceeds $10^8$ above 6 Hz. With additional active shielding the chamber has a shielding factor of more than $7 \cdot 10^6$ down to 0.01 Hz. (Right): Side view of inner chamber (2.9 x 2.9 x 2.9 m$^3$), seen from the door opening. The pneumatically driven sliding door is indicated by a rectangle with thin dashed lines. The big black rectangle is the Dewar, which houses 19 identical modules. Each module is equipped with 16 low-$T_c$ SQUID magnetometers. The big open rectangles are the $B_x$- and $B_y$-coil pairs. The small circle below the Dewar shows the sample cell (without fixation). The (-x)-axis of the chosen coordinate system points at an angle of $28^0$ to the north-south direction (see also Section 4.2).

**Table 1:** Measured field gradients around the sample position with and without magnetic guiding field.

| Gradient | Residual field gradient [pT/cm] | Field gradient [pT/cm] with holding field ($B_x=0.4\mu T$) |
|---|---|---|
| $\partial B_{1,x}/\partial x$ | 2.4 | -27 ÷ -34 |
| $\partial B_{1,x}/\partial y$ | 0.3 | -12 |
| $\partial B_{1,x}/\partial z$ | -1.2 | -4 |
| $\partial B_{1,y}/\partial x$ | 0.6 | -3.3 ÷ 17 |
| $\partial B_{1,y}/\partial y$ | -2.6 | -15 |
| $\partial B_{1,y}/\partial z$ | -1.5 | -33 ÷ -50 |
| $\partial B_{1,z}/\partial x$ | -1.2 | 3÷10 |
| $\partial B_{1,z}/\partial y$ | -2.5 | 10 |
| $\partial B_{1,z}/\partial z$ | -0.6 | -24 |

## 3. Study of the free precession of $^3$He spins in a spherical sample cell.

A sealed-off spherical glass cell of radius $R = 3$ cm filled with $^3$He at $p \approx 4.5$ mbar was put directly beneath one of the central z-plane SQUIDs. The distance $d$ from the centre of the cell to the monitoring SQUID was $d \approx 6$ cm. The longitudinal relaxation time $T_1$ of the cell made from low-relaxation GE180 glass [28-30] had been measured before to be $T_1 = 85 \pm 5$ h in a conventional NMR setup. The gas was optically pumped by means of a 2W Yb-doped fibre laser ($\lambda=1083$ nm) and the polarization build-up along the x-axis could be monitored optically by analysing the circular polarization of the 668 nm fluorescence line of the weak discharge spectrum maintained during the optical pumping process [31].

The polarization obtained was about 15% at that pressure. After switching off the discharge and laser light, a slow (adiabatic) rotation of the magnetic field direction into the y-direction, followed by a fast (non-adiabatic) switch of the magnetic field back into the x direction causes the spins first to orient themselves into the y direction, and then to start to precess in the yz-plane. This procedure is equivalent to a "π/2-pulse". Figure 3a shows the recorded SQUID signal over a time interval of 0.5s at the beginning of the precession cycle. The signal amplitude reaches $A = \Delta B_s =$ 12.5 pT and the precession frequency is $f \approx 13$ Hz. In Fig.3b the exponential decay of the signal amplitude (envelope) over a period of about 10 h is shown. Hence, we can deduce a transverse relaxation time of $T_2^* = (60.2 \pm 0.1)[h]$. To our knowledge, this is the longest spin-coherent relaxation time of a macroscopic sample measured so far. The expected signal amplitude can be calculated from equations (8) and (9) and gives $A_{cal} = \Delta B_{s, cal} \approx 16$ pT. Considering the uncertainties of our input parameters ($^3$He-polarization $P$, $^3$He-pressure $p$, and distance $d$), this result is in fairly good agreement with what has been measured.

The long free spin-precession time $T_2^*$ is also predicted from equation (7): With $D_{He} = 470$ cm$^2$/s at $p = 4.5$ mbar, $B_0 = 400$ nT, and further $|\vec{\nabla} B_{1,x}| = (32.7 \pm 3.5) \, pT/cm$, $|\vec{\nabla} B_{1,y}| = (44.4 \pm 8) \, pT/cm$, $|\vec{\nabla} B_{1,z}| = (27 \pm 0.8) \, pT/cm$ for the magnetic field gradients (Table 1), we expect a $T_{2,field}$ of $T_{2,field} = (370 \pm 64) h$. This result, together with the measured longitudinal relaxation time $T_1$ of the sample cell used, results in a $T_2^*$-time of $T_2^* = (69 \pm 4) h$. This is very close to the measured $T_2^*$, which shows that the main sources of the transverse spin-relaxation are understood quantitatively. This result also demonstrates that even longer spin coherence times of macroscopic samples can be obtained, and that our present value for $T_2^*$ is mainly limited by the wall relaxation time $T_{1,wall}$ of the sample cell.

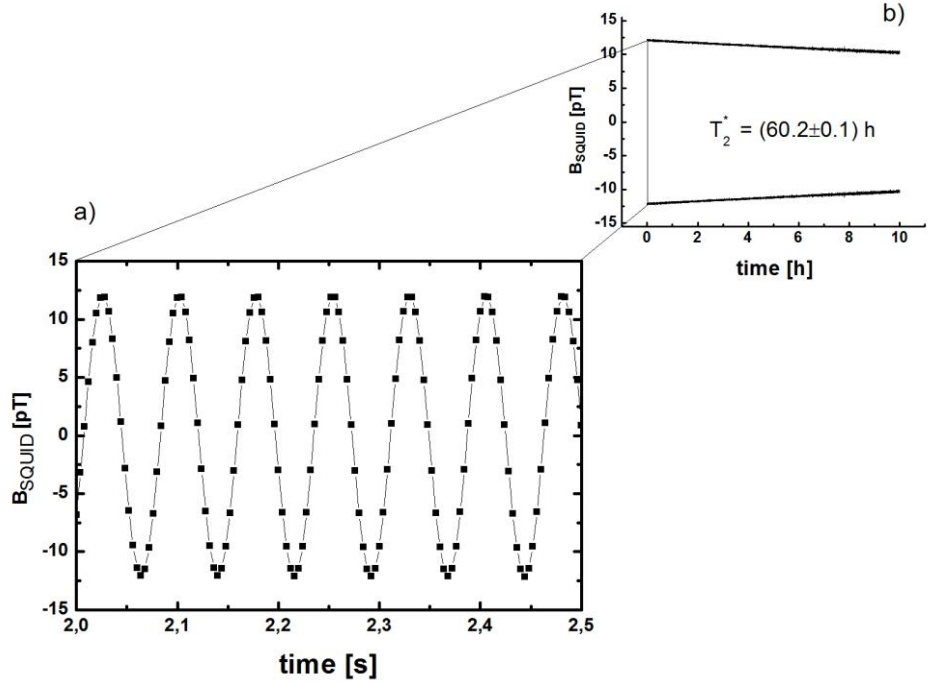

Fig.3 : a) Free spin-precession signal of a polarized $^3$He sample cell recorded by means of a low-$T_c$ SQUID ( sampling rate: 250 Hz). b) Envelope of the decaying signal amplitude. From an exponential fit to the data, a transverse relaxation time of $T_2^* = (60.2 \pm 0.1)[h]$ can be deduced.

Finally, with the measured value of the $SNR(f_{BW} = 1Hz)$ at $T = 0\ s$ being $SNR = 12500 : 2.3 \approx 5400$ (see Fig.1 and Fig.3) we obtain a measurement sensitivity $\delta B$ of

$$\delta B[fT] \approx 3150 \cdot \frac{\sqrt{C}}{T^{3/2}} \ . \qquad (10)$$

Here, we used equation (5) with $\gamma_{He} = 3.24 \cdot 10^7 [Hz/T]$ and took into account the effect of the exponential damping $(\sqrt{C})$ of the free precession signal as derived in the appendix (Eq. A.12). Figure 4 shows the increase in measurement sensitivity $\delta B$ as a function of the observation time $T$ taking $T_2^* = 60\ h$ for the transverse spin relaxation time. The level of $\delta B \approx 1 fT$ is reached after an integration time of $T \approx 220s$ and, according to the $T^{-3/2}$ power law, a measurement sensitivity of $\delta B \approx 1.5 \times 10^{-4}\ fT$ can be reached after one day.

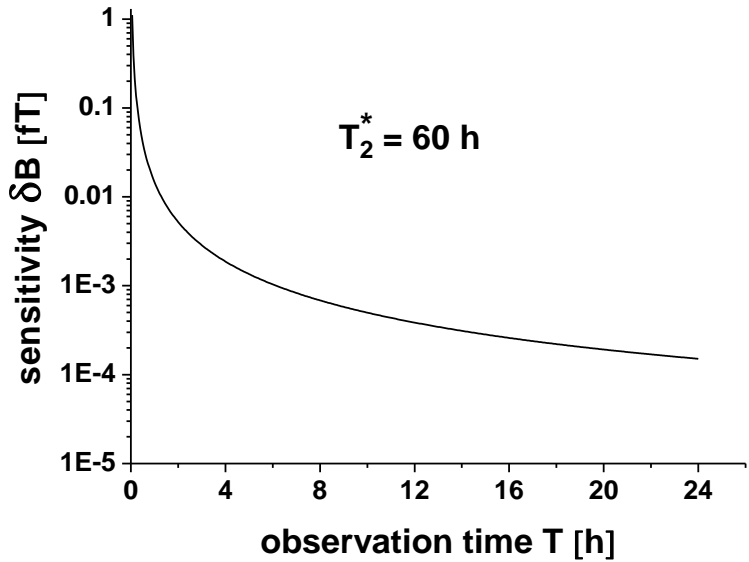

Fig.4: Measured sensitivity (CRLB) in tracing tiny magnetic field fluctuations as a function of the observation time $T$. The expected sensitivity of $1.5 \times 10^{-4}$ fT after one day is not yet limited by the uncertainty of a frequency standard which provides a relative stability of $\approx 10^{-14}$ ($\delta B_{clock} \approx 0.4 \mu T \times 10^{-14} = 4 \times 10^{-6} \, fT$) that minimizes possible sampling rate jitter and drifts below this sensitivity limit.
.

## 4. Applications

In the following we present two applications of our ultra-sensitive magnetometer based on free precession of nuclear spins of $^3$He and $^{129}$Xe. Both applications need the most precise measurement of the magnetic field with long-term stability. The first takes part in the research project of the planned measurement of the electric dipole moment of the neutron (nEDM) at the Paul Scherrer Institut (PSI / Villigen), where the magnetic field has to be measured with highest possible resolution over a period of 1000 s. Secondly, we propose a free spin-precession atomic clock experiment, where both gases ($^{129}$Xe and $^3$He) are in the same cell. Thus, we search for a Lorentz-violation signature by monitoring the relative Larmor frequencies or phases of the co-located $^3$He and $^{129}$Xe as the laboratory reference frame[2] rotates with respect to distant stars. The periods to compare are in the range of one day in this application.

## 4.1 $^3$He magnetometer for neutron EDM measurements

As the limit of the electric dipole moment (EDM) of the neutron is pushed lower, a principal source of error becomes variations of the magnetic field in the spectrometer region, especially

---

[2] quantization axis of the experiment directed under an angle of $28^0$ east-west in the Earth's reference frame, see Fig.2

those from leakage currents that are systematically coupled to the direction of the applied electric field. In this respect an ideal monitor of magnetic field fluctuations would be a magnetometer with a sensitivity of less than 10 fT, in order not to limit (due to normalization) the statistical accuracy or its equivalent in terms of magnetic field fluctuations of $\delta B \approx 50$ fT, which is expected for the planned neutron EDM experiment at PSI during one Ramsey spin precession cycle of 200s, typically [32].

The cohabiting Hg-magnetometer used before at ILL [33] has reached its limits of sensitivity ($\delta B \approx 200$ fT). Furthermore, due to a geometric phase effect [34], particles trapped in electric and magnetic fields accumulate a phase shift, which is linear in $E$ and thus cannot be distinguished from a true EDM effect. As a consequence, the magnetometer has to operate in an $E$-field free region and thus in a separate volume than the ultra cold neutrons (UCN) unless interparticle-collisions suppress these correlated effects [35].

The use of $^3$He as a magnetometer has already been proposed in 1984 by Ramsey [36]. In this proposal, $^3$He would have been used as a cohabiting magnetometer, like Hg, which was later used for that purpose. In the following, we show the layout of our neutron spectrometer, test measurements with a prototype of a flat magnetometer vessel with $^3$He, and we present Monte-Carlo simulations to estimate the temporal response of such a large magnetometer vessel.

## Layout of a neutron EDM spectrometer

In our concept, UCN and $^3$He will be in separate volumes and the general layout of the proposed neutron EDM spectrometer at PSI is sketched in Fig.5a. The whole setup sits within a multi-layer µ-metal shield (not shown) and is immersed in a weak uniform magnetic guiding field $B_z$ of typically 1 µT pointing in vertical direction. Spin-polarized $^3$He gas enters through an inlet valve and expands into two flat cylindrical magnetometer vessels sandwiching the double-chambers for UCN storage. Each magnetometer vessel will have a volume of about 9 litres, assuming an inner diameter of $\Phi_{in} = 54$ cm and height $L_{in} = 4$ cm, and a $^3$He gas pressures of about 1 mbar. After filling, a π/2-pulse is applied, which causes free precession of the $^3$He spins around the static $B_z$-field. The $^3$He spin precession will be monitored directly by means of SQUIDs or, alternatively, by use of Cs-magnetometers, which are placed close to the sidewalls at angles of $120^0$.

The sandwich-type of arrangement (**t**op, **b**ottom) of the flat cylindrical magnetometer vessels cover the same magnetic flux than the UCN double-chamber to a good approximation. In the UCN chamber, the spin-precession frequency of UCN ($\overline{\omega}_{UCN}$) is measured by the Ramsey technique of separated oscillating fields [37]. Hence, the average value of frequency measurements $\overline{\omega}_{He} = (\overline{\omega}_{t,He} + \overline{\omega}_{b,He})/2$ gives the normalization signal for the UCN free precession frequency $\overline{\omega}_{UCN}$, whereas the frequency difference determines the magnetic field gradient $\langle \partial B / \partial z \rangle = (\overline{\omega}_{t,He} - \overline{\omega}_{b,He})/(\gamma \cdot \Delta z)$ to a high precision with Δz being the distance between the upper and lower magnetometer vessels.

## Prototype of a flat 3He magnetometer

At the PTB-Berlin, test measurements have been performed at BMSR-1 with prototypes of flat cylindrical magnetometer vessels ($R_{in} = 14$ cm, $L_{in} = 5$ cm) made from hardened Borosilicate glass, which were filled with $^3$He at a pressure of around 1 mbar. The vessels were optically pumped ($P \approx 15\%$) along their cylinder axis (see Fig.5b) aligned parallel to the magnetic guiding field $B_x$ ($B_x \approx 400$ nT) [3]. After applying a "π/2-pulse", the free precession of the

---
[3] same arrangement of coils as in BMSR-2 (see Fig.2)

$^3$He spins in the yz-plane was monitored by means of a vertical (z-direction) one-channel low-T$_c$ DC-SQUID (white system noise level $\rho_{System} \approx 4.5$ fT/$\sqrt{Hz}$) which was positioned close to the sidewalls of the glass vessel(s). The magnetic field amplitude of the precessing spins is expected to be $\Delta B_{He} = 13.5$ pT at a radial distance of 17.5 cm from the centre (SQUID position) for a $^3$He polarization of 15% (see Fig.5c). Due to the uncertainties of exact SQUID positioning, $^3$He-pressure and -polarization, we estimated an overall error of $\pm 2$pT for $\Delta B_{He}$. Theoretical expressions for the magnetic field gradient induced transverse relaxation rate of a spin-polarized gas confined in a cylindrical sample cell were derived by McGregor [18]. In the regime of motional narrowing this gives

$$\frac{1}{T^*_{2,field}} \sim \frac{\gamma^2 L^4}{120 \cdot D}\left(\frac{\partial B_x}{\partial y}\right)^2 + \frac{7\gamma^2 R^4}{96 D}\left(\frac{\partial B_x}{\partial z}\right)^2 \qquad (11)$$

Like in spherical sample cells (Eq.(7)), $1/T^*_{2,field}$ scales with $1/T^*_{2,field} \propto (length)^4 \cdot pressure$. Therefore, we expect for our large cylindrical magnetometer vessels a significant increase of the field-gradient induced relaxation rate of about a factor of 160 as compared with $(1/T^*_{2,field})_{sp} \approx 1/370h$ for the spherical cell of radius 3 cm. Still, a resulting $T^*_{2,field}$ of $T^*_{2,field} \approx 370h/160 \approx 2h$ is long enough to monitor the free spin-precession over a typical Ramsey cycle of 200s.

Figure 6 shows the result for one magnetometer vessel. The measured transverse relaxation time $T^*_2$ is about 60 min, which is the combined effect of field-gradient relaxation and wall relaxation. The latter one has been determined to be $T^{cyl}_{1,wall} = 2.1h$ in a conventional NMR setup. The signal amplitude of 12 pT measured at the beginning of the spin-precession cycle agrees quite well with the calculated $\Delta B_{He} = 13.5(20)$ pT from Fig.5c. Knowing the white noise level of our DC-SQUID inside BMSR-1, we again obtain a *SNR* of *SNR* > 2500 in a bandwidth of 1Hz.
According to equation (5), this results in a measurement sensitivity $\delta B$ of

$$\delta B \approx 2 fT \qquad (12)$$

during one Ramsey cycle (*T=200s*).
Next, we investigated the arrangement of two cylindrical magnetometer vessels at a distance of $\Delta x = 12$ cm, each of them filled with $^3$He at 1 mbar (see Fig.5b). Both vessels were optically pumped simultaneously along the direction of the magnetic guiding field ($B_x$) and, finally, a "π/2-pulse" was applied in order to let the spins precess in the plane perpendicular to the magnetic guiding field. The DC-SQUID which was positioned in between both vessels, as indicated in Fig.5b, monitored a beat-signal as a result of the two interfering spin-precession signals, which is shown in Fig.7a. From the exponential decay of the envelope, the transverse relaxation time $T^*_2$ was determined to be $T^*_2 = (1159 \pm 30)s$. The reduced $T^*_2$ is due to the stronger magnetic field gradients, since in the double cell arrangement both magnetometer vessels had to be positioned somewhat outside the symmetry plane of the $B_x$ field. Figure 7b shows the resulting power spectrum of the spin-precession signal from which the average spin-precession frequency in each of the two vessels can be determined. From the frequency difference $\Delta f = 0.0746\ Hz$, the average field gradient $\langle \partial B_x / \partial x \rangle$ over the sensitive area $A = \pi \cdot R^2 \approx 700\ cm^2$ of the two flat cylindrical vessels can be derived to be $\langle \partial B_x / \partial x \rangle = 2\pi \cdot \Delta f / (\Delta x \cdot \gamma_{He}) = 194\ pT/cm$. This value of the field gradient inside BMSR-1 is in good agreement with values measured independently in ref.[14] from spin precession of $^{129}$Xe in two adjacent spherical bulbs.

## Monte-Carlo simulations for the temporal response of a flat 3He magnetometer

At this point it is worth to think more closely on the temporal response of such a large area $^3$He-magnetometer to a local change of the magnetic field ( for example, induced by a step-like rise of leakage currents at the central HV electrode in between the two UCN storage chambers): In a Monte-Carlo simulation we considered the diffusion of $^3$He spins confined in our flat cylindrical glass vessel and studied the response of the SQUID detector to a sudden change of the magnetic field at its centre. As expected, the system takes a certain time ($\tau$) until it reaches a new equilibrium which reflects the new average magnetic field $B_0 + \langle \Delta B \rangle$ across the sensitive area of the cylinder. The response times strongly depend on the $^3$He gas pressure (diffusion) reaching 50 ms at pressures around 0.5 mbar (see Fig.8). Since the precessing spins of the stored UCN experience a similar delay in their accumulated phase $\langle \Delta \Phi \rangle \approx \gamma_n \cdot \langle \Delta B \rangle \cdot \tau_{n,diff}$ during a Ramsey cycle with $\tau_{n,diff} \approx \tau_{coll} = \lambda / \langle v \rangle \approx 25$ ms [4] residual normalization errors due to different temporal responses can be kept low.

With the later use of 3 SQUID detectors around each magnetometer vessel, as indicated in Fig.5a, the relative phases of the precessing magnetic moment of the sample spins with respect to the SQUIDs are fixed, causing a further reduction of the signal noise in the final read-out.

---

[4] The mean free path λ of UCN in a cylindrical cell is $\lambda = 4V/A = 2RL/(R+L)$. Together with $\langle v \rangle = 2/3 \cdot v_F \approx 330\ cm/s$, the mean collision time $\tau_{coll}$ gets $\tau_{coll} = \lambda / \langle v \rangle$, which is $\tau_{coll} \approx 25\ ms$ for the magnetometer vessels used.

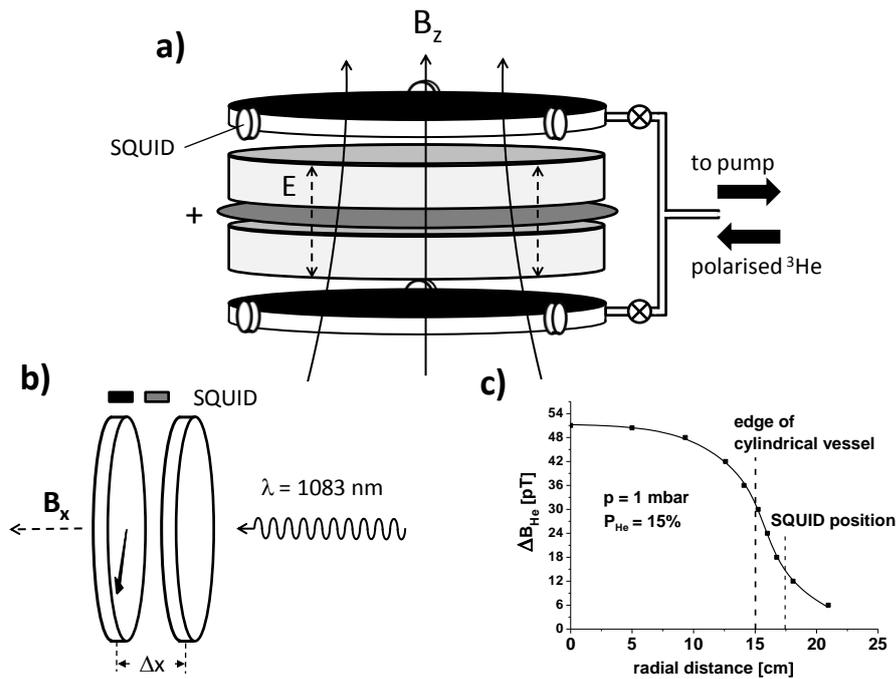

Fig.5: a) Schematic sketch of the proposed UCN spectrometer at PSI with the two $^3$He magnetometer vessels on top and bottom. b) Arrangement of flat cylindrical glass vessels ($R_{in}$=14 cm, $L_{in}$=5cm) at BMSR-1 which are filled with $^3$He at $p$ = 1mbar and optically pumped along the $B_x$ magnetic guiding field ($B_x \approx$ 400 nT); after applying a "π/2-pulse", the free spin precession is monitored by means of a SQUID. Different SQUID positions used if only one magnetometer vessel was activated (black) or the beat signal from both cells was analyzed (grey). c) calculated magnitude of the rotating magnetic field $\Delta B_{He}$ in radial direction.

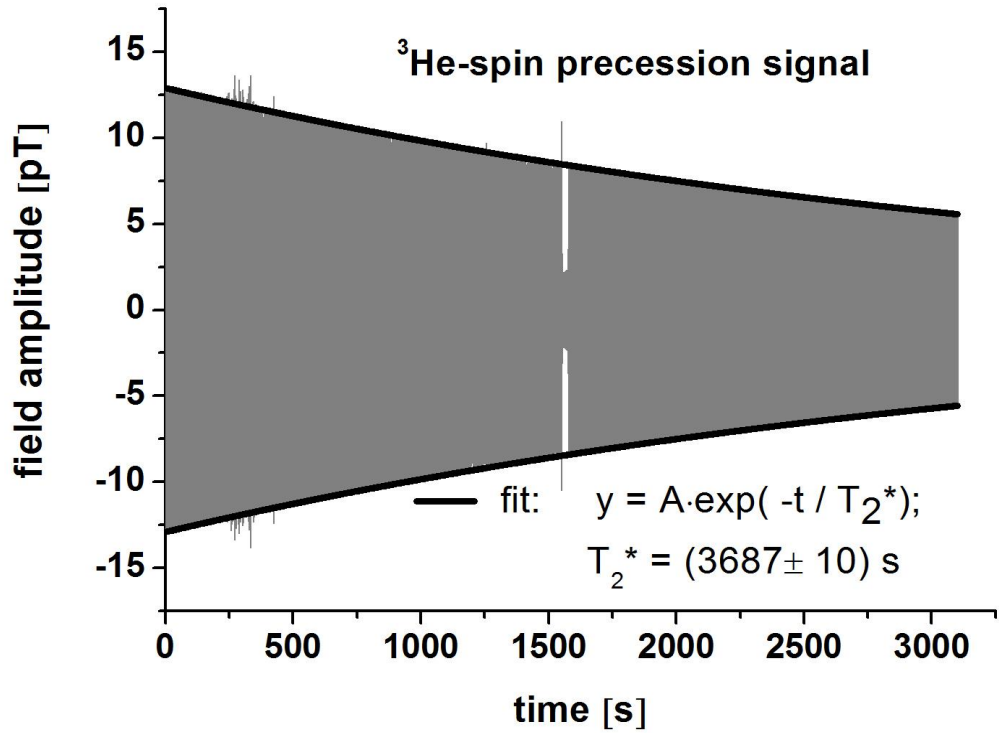

Fig.6: Measured ³He spin-precession signal in a flat cylindrical magnetometer vessel of $R_{in}$ =14 cm and $L_{in}$ = 5 cm by means of a low-$T_c$ SQUID. The characteristic time constant for the decay of the signal amplitude gives $T_2^* = (3687 \pm 2)s$ (fit). External disturbances during data acquisition caused some spikes in the measured signal amplitude at $t \approx$ 300s and $t \approx$ 1600s.

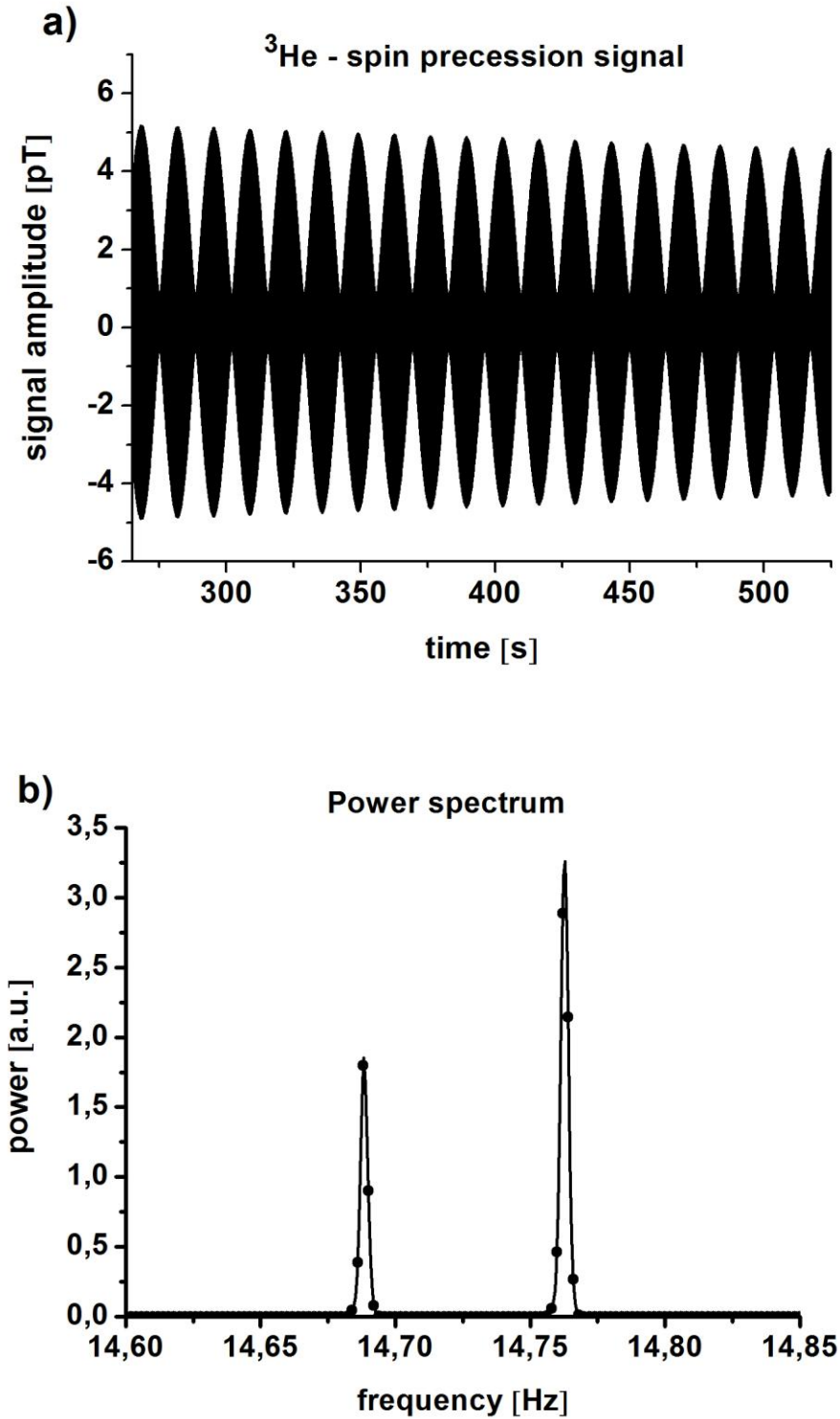

Fig.7: a) Observed beat-signal from $^3$He spin-precession using two flat cylindrical magnetometer vessels at distance $\Delta x$=12 cm with their axes aligned along the field axis (x-axis). The SQUID was positioned in between both vessels at a radial distance of 17.5 cm from the cylinder axis (see Fig.5b). From the exponential decay of the envelope a $T_2^*$ of (1159±30) s is deduced. b) Resulting power spectrum of the spin-precession signal from which the average spin-precession frequency in each of the two vessels could be determined.

The spectral amplitude of the second vessel (the more downstream one with respect to the incoming laser light ) is somewhat lower, due to the smaller polarization obtained there. From the frequency difference $\Delta f$ = 0.0746 Hz the average field gradient $\langle \partial B_x / \partial x \rangle$ was determined to be $\langle \partial B_x / \partial x \rangle = 2\pi \cdot \Delta f /(\Delta x \cdot \gamma_{He}) = 194 \, pT/cm$

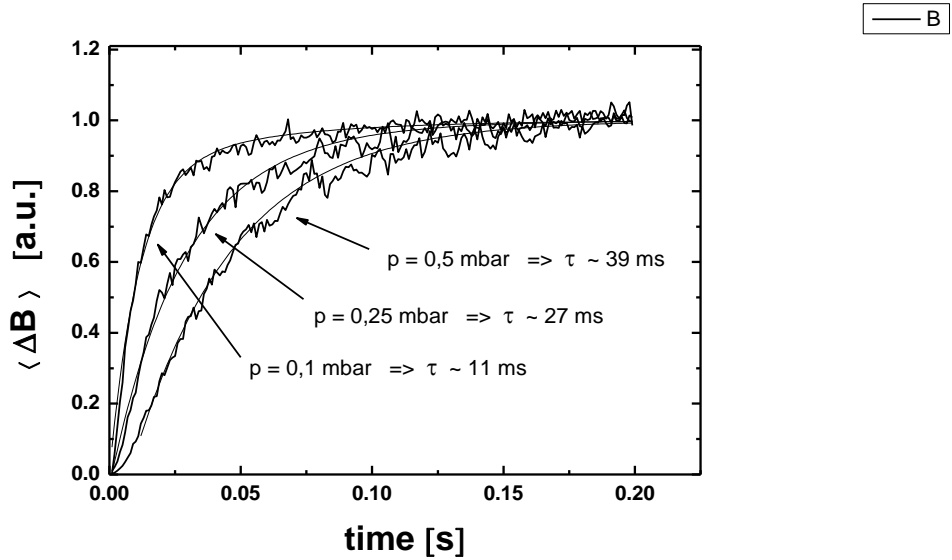

Fig.8: Monte-Carlo simulation: Temporal response of the $^3$He-magnetometer to a local change of the magnetic field at its centre (cylinder vessel). An exponential fit to the MC data gives the characteristic time constants for each gas pressure investigated.

## 4.2. $^3$He/$^{129}$Xe clock comparison experiments

Precision measurement of the Zeeman splitting in a two-state system is important for magnetometry, as well as for searches for physics beyond the standard model [38-44]. The most precise tests of new physics are often realized in differential experiments which compare the transition frequencies of two co-located clocks, typically radiating on their Zeeman or hyperfine transitions. The advantage of differential measurements is that they render the experiment insensitive to common systematic effects, such as uniform magnetic field fluctuations [43]. Lorentz symmetry is a fundamental feature of modern descriptions of nature, including both the standard model of particle physics and general relativity. However, both theories are believed to be the low-energy limit of a single fundamental theory at the Planck scale. Even if the underlying theory is Lorentz invariant, spontaneous symmetry breaking might result in small apparent violations of Lorentz invariance at an observable level [45]. Experimental investigations of the validity of Lorentz symmetry therefore provide valuable tests of the framework of modern theoretical physics. Kostelecky and co-workers have developed a standard model extension that treats the effects of spontaneous Lorentz symmetry breaking in the context of a low-energy effective theory, in which terms can be induced which appear to violate Lorentz invariance explicitly [46].
We propose a free spin-precession $^3$He/ $^{129}$Xe atomic clock experiment to search for a Lorentz-violating signature by monitoring the relative Larmor frequencies or phases of the co-located $^3$He and $^{129}$Xe as the laboratory reference frame rotates with respect to distant stars.

More precisely, we are searching for a sidereal variation of a combination of Larmor frequencies of the form

$$\Delta \upsilon = \upsilon_{He} - \frac{\gamma_{He}}{\gamma_{Xe}} \cdot \upsilon_{Xe} = \delta\upsilon_x \cdot \cos(\Omega_s \cdot t) + \delta\upsilon_y \cdot \sin(\Omega_s \cdot t),  \tag{13}$$

where $\Omega_s$ is the angular frequency of the sidereal day ($\Omega_s/2\pi$ = 1/ 23,934h), and the parameters ($\delta\upsilon_x, \delta\upsilon_y$) represent the net effect of Lorentz-violating couplings on the $^3$He/$^{129}$Xe frequency. In this combination of frequencies, the sensitivity to magnetic field fluctuations cancels to the first order. To date, co-located $^{129}$Xe and $^3$He Zeeman masers [47] set the most stringent limit on leading order Lorentz-violation, consisted with no effect at the level of $|h \cdot \Delta \nu| \leq 10^{-31} GeV$ as derived from a frequency variation of

$\Delta \nu = \sqrt{\delta\nu_x^2 + \delta\nu_y^2} = (53 \pm 45) nHz$ (67% C.L.) for the $^3$He maser frequency with the $^{129}$Xe maser acting as a co-magnetometer to stabilize the systems static magnetic field.

In Section 3 we have shown that a free spin-precession $^3$He magnetometer based on SQUID detection can reach a measurement sensitivity of $\approx$1fT after about 4 min. Using equations (3) and (4) this translates into a 1$\sigma$-level for the $^3$He frequency estimation of $\sigma_\nu$ < 0.01 nHz (CRLB) after a data taking time of just one day. Frequency instabilities should be even smaller than they were in the maser experiment, since there are no obvious couplings to external sources as it was the case for the laser driven $^3$He/$^{129}$Xe Zeeman maser (e.g., noble-gas polarization induced frequency shift, etc). Unfortunately, spin-polarized $^{129}$Xe does not have such good relaxation properties than $^3$He. The present size of its longitudinal relaxation ($T_{1,Xe}$) limits the transverse relaxation times ($T_{2,Xe}^*$) to be a few hours. As discussed in detail in ref. [48], the $T_{1,Xe}$-time of $^{129}$Xe in our pressure range is the combined effect of the density independent wall relaxation $T_{1,wall}$ and the $^{129}$Xe-$^{129}$Xe molecular spin relaxation $T_{1,vdW}$ given by

$$\frac{1}{T_{1,Xe}} = \frac{1}{T_{1,wall}} + \frac{1}{T_{1,vdW}} \tag{14}$$

with $1/T_{1,vdW} = \Gamma_{vdW}^{Xe}/(1 + r \cdot [X]/[Xe])$. $\Gamma_{vdW}^{Xe} = 1/4.1h$ is the pure Xenon rate and $r$ =1.05 is the relative breakup coefficient for the van der Waals molecules by introducing a buffer gas such as N$_2$ at density [N$_2$]. Thus, for high buffer gas ratios [N$_2$]/[Xe] the molecular relaxation is shortened and one is left with $T_{1,Xe} \approx T_{1,wall}$. Then, our problem is reduced to finding a low-relaxation container for the polarized Xenon gas. Since both $T_{2,Xe}^*$ and $T_{2,He}^*$ have to be optimized (Eq.(6)), the influence of the increased total gas pressure on the field gradient induced transverse relaxation time $T_{2,field}$ has to be reconsidered: In a gas mixture (GM) with N$_2$ as buffer gas, the resulting diffusion coefficients for $^3$He and $^{129}$Xe are given by [49]

$$\frac{1}{D_{He}^{GM}} = \left(\frac{p_{He}}{D_{He}} + \frac{p_{Xe}}{D_{He\,in\,Xe}} + \frac{p_{N2}}{D_{He\,in\,N2}}\right) \text{ and } \frac{1}{D_{Xe}^{GM}} = \left(\frac{p_{Xe}}{D_{Xe}} + \frac{p_{He}}{D_{Xe\,in\,He}} + \frac{p_{N2}}{D_{Xe\,in\,N2}}\right) \tag{15}$$

with $D_{He\,in\,Xe} \approx 600$ cm$^2$/s [50], $D_{He\,in\,N2} \approx 770$ cm$^2$/s [50], $D_{Xe\,in\,He} \approx 790$ cm$^2$/s [51], and $D_{Xe\,in\,N2} \approx 210$ cm$^2$/s [51], where the partial pressures are given in units of mbar.

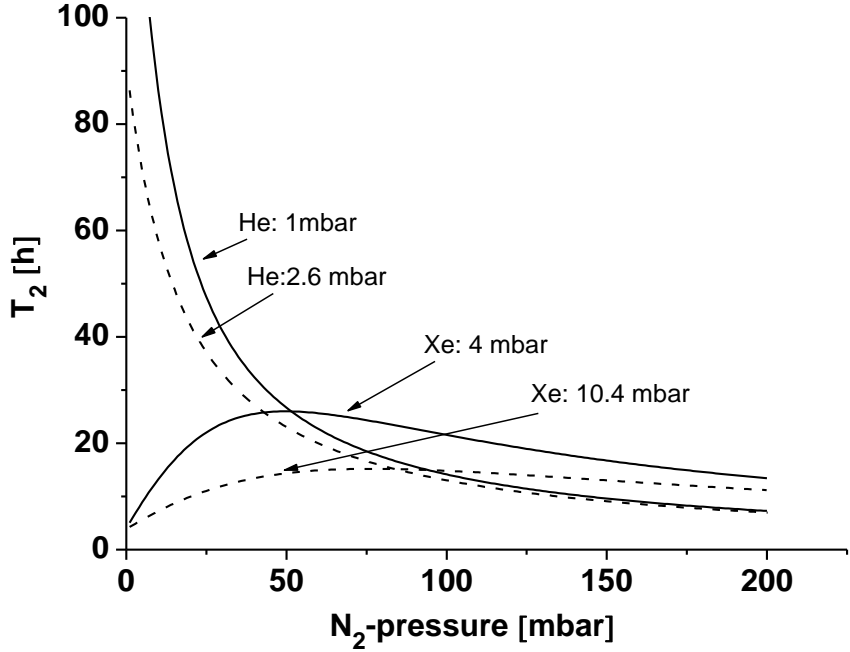

Fig.9: Calculated effect of magnetic gradient and molecular spin relaxation as a function of the $N_2$ buffer gas pressure in $^{129}$Xe/$^3$He gas mixtures with [$^{129}$Xe]/[$^3$He] = 4:1.

In Fig.9, the dependence of $T_2$ with $1/T_2 = 1/T_{2,field} + 1/T_{1,vdW}$ is plotted as a function of the $N_2$ partial pressure using the same field gradients as listed in Table 1. For $^3$He, the formation of van der Waals molecules can be neglected, i.e., $1/T_{1,vdW} = 0$, and we see the decrease of $T_{2,field}^{He}$ as the total pressure is increased. In case of $^{129}$Xe, we find maxima in $T_2^{Xe}$ at $N_2$ partial pressures around 50 mbar and 100 mbar for a fixed ratio of [Xe]/[He] =4:1 at $p_{He}$ = 1mbar and $p_{He}$ = 2.6 mbar, respectively. The ratio [Xe]/[He] =4:1 was chosen in order to obtain the same magnetzation for both gas species (assuming an equal degree of nuclear polarization), since $\gamma_{He}/\gamma_{Xe} \approx 2.75$ and while the $^3$He gas is pure, the $^{129}$Xe is isotopically enriched to about 70%.

We used in our first test measurement a gas mixture with partial pressures of $p_{He}$=2.6 mbar, $p_{Xe}$=10.4 mbar, and $p_{N2}$=62 mbar, and obtained $T_2^{He} \approx 19.5h$ and $T_2^{Xe} \approx 15h$ (see Fig.9). Hence, taking into account the measured wall relaxation times of $T_{1,wall}^{He} \approx 80h$ and $T_{1,wall}^{Xe} \approx 5h$ in the spherical GE180 glass vessel[5] used, we expect for both noble gases a transversal relaxation time of $T_{2,He}^* \approx 15,6h$ and $T_{2,Xe}^* \approx 3.7h$, respectively.

The sample cell was filled consecutively with $^3$He, $^{129}$Xe, and $N_2$ using a manifold connected to the three adjacent gas supplier cells via a regulating valve. In Fig.10 the measured signal amplitude of the precessing co-located $^3$He/$^{129}$Xe spins is shown using the same arrangement as described in Section 2. The extracted transverse relaxation times for $^3$He and $^{129}$Xe ($T_{2,He}^* \approx 17h$, $T_{2,Xe}^* \approx 3.3h$) are in good agreement with the predicted numbers.

---

[5] a cell with a radius of 3 cm, an appendix leading to a stopcock valve and a glass flange

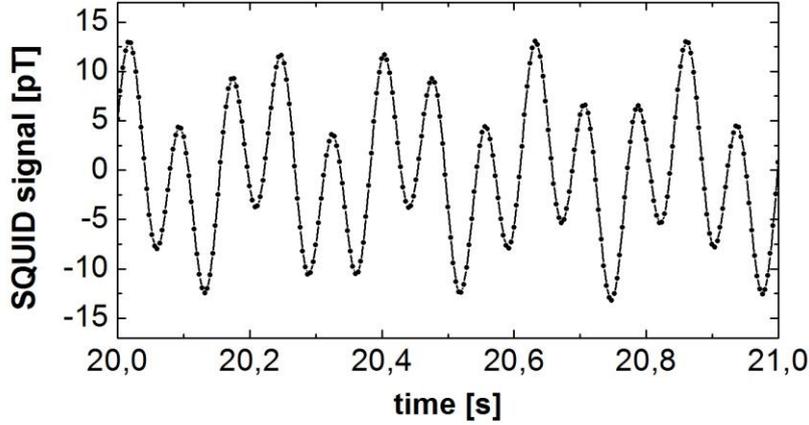

Fig.10: direct SQUID readout of the co-precessing $^3$He/$^{129}$Xe spins.

Data processing in order to extract the $^3$He/$^{129}$Xe phases and amplitudes is performed as follows: First, the measured SQUID signal $s(t)$ is mixed numerically with a reference frequency[6] $\overline{\omega}_{He(Xe)}$ according to $s(t) \cdot \exp(-i\overline{\omega}_{He(Xe)} \cdot t)$ and is then transformed into the frequency domain via direct Fourier transformation (FFT). After that, an exponential filter $\sim \exp\left(-(\omega/\omega_{cut})^2\right)$ is applied. Its cut-off frequency $\omega_{cut}$ determines the bandwidth of our output data. The filtered data are then transformed back into the time domain using inverse FFT. The result is $\mathcal{F}_{He(Xe)}(t)$. The phase $\Phi_{He(Xe)}(t)$ is found as

$$\Phi_{He(Xe)}(t) = atan(\mathrm{Im}[\mathcal{F}_{He(Xe)}(t)]/\mathrm{Re}[\mathcal{F}_{He(Xe)}(t)]), \tag{16}$$

and the amplitude is $|\mathcal{F}_{He(Xe)}(t)|$. We take the weighted difference between the He- and the Xe phases motivated above,

$$\Delta\Phi(t) = \Phi_{He}(t) - \gamma_{He}/\gamma_{Xe} \cdot \Phi_{Xe}(t) \tag{17}$$

with $\gamma_{He}/\gamma_{Xe} = 2.75408159(20)$ (using the literature values of their gyromagnetic ratios [52,53]). The Zeeman term from the applied magnetic field and with it the temporal field fluctuations should drop out and one is left with a possible sidereal modulation of the phase difference which, according to equation (13), is given by

$$\Delta\Phi_{LV}(t) = 2\pi/\Omega_s \cdot \left(\delta\upsilon_x \cdot \sin(\Omega_s \cdot t) - \delta\upsilon_y \cdot \cos(\Omega_s \cdot t)\right) \tag{18}$$

---

[6] As reference frequencies $\overline{\omega}_{He}$ and $\overline{\omega}_{Xe}$ of the particular data acquisition cycle, we take the mean spin-precession frequencies of $^3$He and $^{129}$Xe as determined from the power spectral density obtained by FFT.

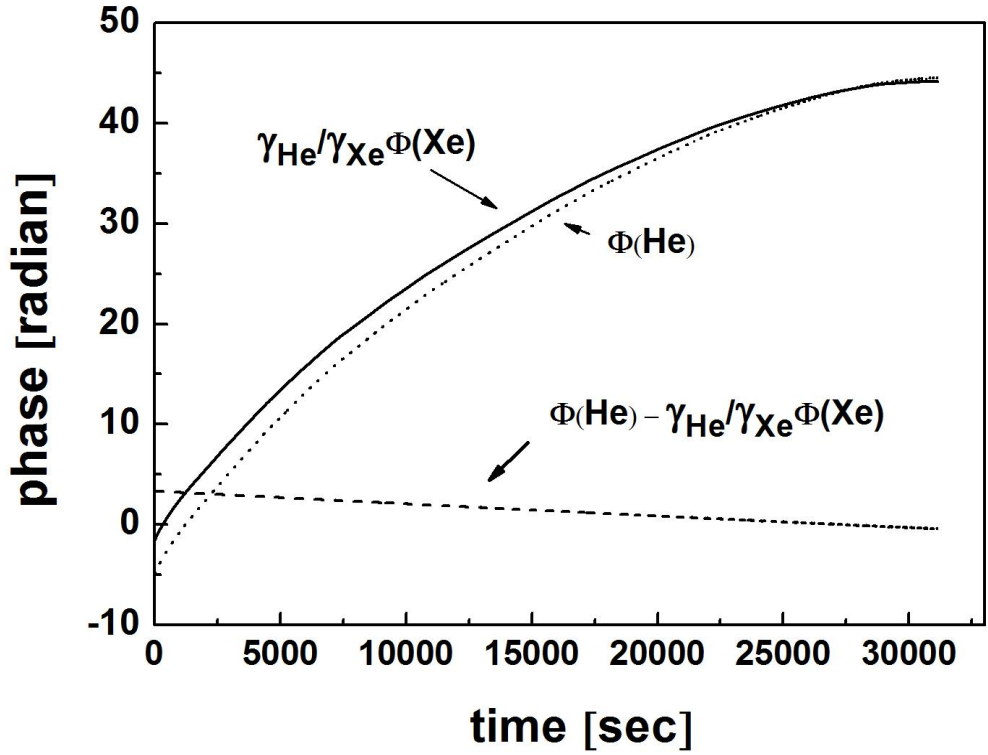

Fig.11: Extracted phase signal $\Phi_{He}(t)$ and $\Phi_{Xe}(t)$ from the co-precessing $^3$He and $^{129}$Xe sample spins. The phase difference $\Delta\Phi(t) = \Phi_{He}(t) - \gamma_{He}/\gamma_{Xe} \cdot \Phi_{Xe}$ should be sensitive to possible LV-terms described by equation (18).

Figure (11) shows the temporal change of the phase difference $\Delta\Phi(t)$ measured in a long run[7] with a total acquisition time of T=31072 s. Besides a general phase offset, we found an almost linear decrease of $\Delta\Phi$. The linear dependence of $\Delta\Phi$ could be caused by a possible chemical shift, e.g., due to adsorption at the walls of the glass bulb, which affects Xe stronger than He and which may lead to a deviation of the ratio of their respective gyromagnetic ratios. Also the fact, that the centres of gravity of the light ($^3$He) and heavy gas ($^{129}$Xe) do not coincide, gives rise to a frequency shift and thus to a linear phase shift in presence of a finite magnetic field gradient. A more detailed analysis of these effects will be given in a forthcoming paper.

Since the primary focus of this paper lies on the analysis of measurement sensitivity for such a $^3$He/$^{129}$Xe clock comparison experiment, $\Delta\Phi(t)$ was fitted with a 2$^{nd}$ order polynomial ($\Delta\Omega_{fit} = a + b \cdot t + c \cdot t^2$) in order to extract the phase residuals. In Fig.12 the phase noise evolution is shown after the subtraction of such a polynomial fit. Due to the exponential decrease of the $^3$He/$^{129}$Xe signal amplitudes (in particular the Xenon amplitude, which decays with a measured relaxation time of $T_{2,Xe}^* = 2.31(1)h$), the residual phase noise rises in time. The inset of Fig.12 shows the time evolution of the root mean square (RMS) together with an exponential fit given by

$$\sigma_{\Phi res}[mrad] = G \cdot \exp(t/T_x) \quad . \tag{19}$$

---

[7] In this run, the magnetic field gradients at the position of the sample cell were about a factor of 2 worse, causing a reduced $T_2^*$.

The data was best described with $G = 0.718(42)\,mrad$ and $T_x = 2.39(5)h$. The fact, that we almost exactly observe $T_x \approx T_{2,Xe}^*$ (within the 2σ-error) can be explained by use of equation (3) and the statistical error propagation law, showing that the phase noise scales like

$$\sigma_{\Phi res} \propto \exp(t/T_{2,Xe}^*) \cdot \sqrt{\exp\left(-2t \cdot \left(\frac{1}{T_{2,Xe}^*} - \frac{1}{T_{2,He}^*}\right)\right) + (\gamma_{He}/\gamma_{Xe})^2 \cdot (A_{He}/A_{Xe})_{t=0}^2} \tag{20}$$

With $(A_{He}/A_{Xe})_{t=0} = 1.802$ for the ratio of the measured signal amplitudes at t=0 and $T_{2,He}^* = 7.94(2)h$, the time dependence of equation (20) is almost entirely determined by its first term, i.e., $\sigma_{\Phi res} \propto \exp(t/T_{2,Xe}^*)$.

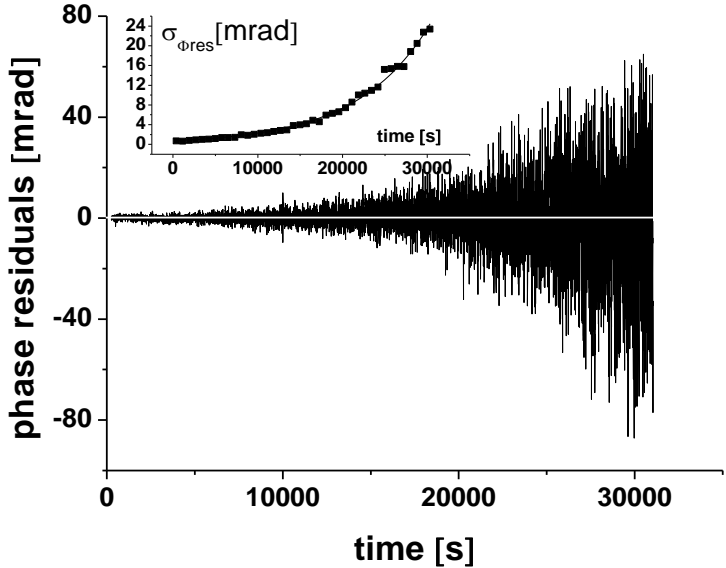

Fig.12: Phase residuals (bandwidth: $f_{BW} = 0.125$ Hz) after subtraction of polynomial fit $\Delta\Phi_{fit} = 3.339528(41) - 1.29902(13) \cdot 10^{-4} \cdot t + 3.0770(78) \cdot 10^{-10} \cdot t^2$ from the data. The white solid line is a fit to the phase residuals using equation (18) with $\delta v_x$ and $\delta v_y$ as free parameters (see text). Inset: Increase of the RMS of the phase noise with time together with an exponential fit.

As mentioned already in the introduction, the Allan Standard Deviation (ASD) plot $\sigma_{ASD}(\tau)$ is a graphical data analysis tool made for the examination of the low-frequency component of time series ($\tau$) and to identify the power-law model for the phase-noise spectrum under study [9]. Therefore, a double logarithmic plot of the dependence of $\sigma_{ASD}$ on $\tau$ is a valuable tool for assigning the origin of the noise processes that may limit the performance of our co-located $^3$He/$^{129}$Xe atomic clock. For white noise, $\sigma_{ASD}$ coincides with the classical standard deviation and we expect a $\sigma_{ASD} \sim \tau^{-1/2}$ dependence on the integration time $\tau$. Figure (13) shows the ASD of the residual phase noise which indeed decreases $\sim \tau^{-1/2}$. This result demonstrates the quality of eliminating the sensitivity to magnetic field fluctuations - at least for our measurement interval of T ≈ 31000 s - which would otherwise lead to a different phase-noise spectrum in the ASD plot. The result also implies that possible noise sources inherent to the $^3$He- or $^{129}$Xe magnetometer don't show up in the ASD plot and thus are not limiting the measured sensitivity (CRLB) as shown in Fig.4.

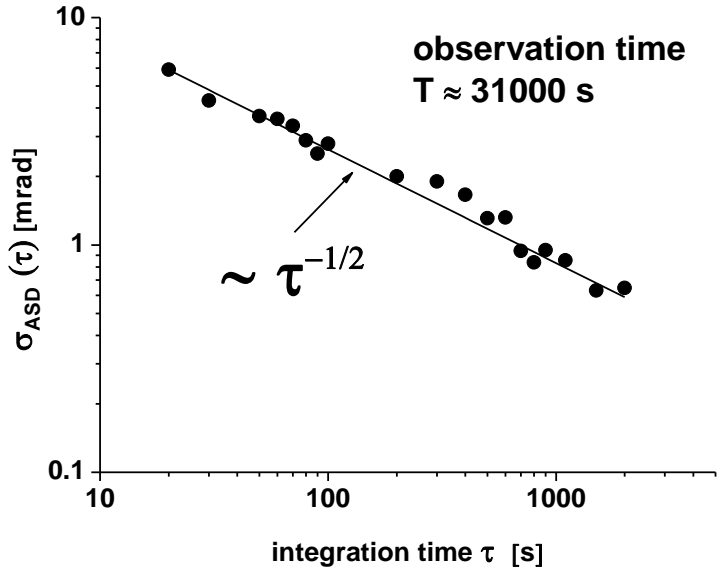

Fig.13: Allan standard deviation (ASD) of the residual phase noise measured with the co-precessing $^3$He/$^{129}$Xe sample spins. For integration times above 4 s ($f_{BW} = 0.125$ Hz) the observed fluctuations decrease as $\tau^{-1/2}$ indicating the presence of a white phase noise amplitude. To fulfil the ASD statistics criteria (N-1) >> 1, we only show data for $\tau \leq 2000$s where we have N-1 ≥ 14 with $N = T/\tau$.

The measurement sensitivity on a possible Lorentz-violating sidereal modulation of the residual phase plot can be obtained from a fit of equation (18) to the data resulting in

$$\delta v_x = (0.35 \pm 2.95) \cdot 10^{-9} Hz \text{ and } \delta v_y = (0.77 \pm 3.70) \cdot 10^{-9} Hz \qquad (21)$$

for the amplitude of Lorentz-violating couplings on the $^3$He/$^{129}$Xe frequency. This fit is shown in Fig.12, too.

In principle, this result can be used to derive new upper limits on leading order Lorentz-violations of the neutron[8], giving $\Delta v = \sqrt{\delta v_x^2 + \delta v_y^2} = (0.8 \pm 4.7) \, nHz$ (67% C.L.) or

$$|\Delta \upsilon \cdot h| \leq 1.9 \cdot 10^{-32} \, GeV \tag{22}$$

---

[8] In case of $^3$He and $^{129}$Xe, the neutron is determining the ground-state properties of the nucleus according to the Schmidt model [46].

However, care has to be taken, since we only recorded a small section (T≈31000s) of a possible sidereal variation of the relative $^3$He/$^{129}$Xe frequency. A finite violation of these fundamental symmetries parametrized by equation (18) can therefore be masked by the linear and quadratic term of the 2$^{nd}$ order polynomial fit used to extract the phase residuals from the measured phase difference (Eq.(17)) .
Still, the result clearly demonstrates the potential of a co-located $^3$He/$^{129}$Xe atomic clock based on free spin precession which to first order is free of magnetic field drifts and which can reach a measurement sensitivity of < 1 nHz.

## 5. Conclusion and outlook

We have presented an ultra-sensitive $^3$He ($^{129}$Xe) magnetometer based on detection of the free nuclear spin precession with a SQUID as low-noise magnetic flux detector. The characteristic spin precession time can be as long as several days in low magnetic fields ( ≈ 1μT) and in the regime of motional narrowing. For observation times $T \approx 200$s, the sensitivity of this magnetometer reaches ≈ 1 fT and, according to the CRLB power law ~ $T^{-3/2}$, it approaches the ≈ $10^{-4}$ fT level (100 Zeptotesla) after one day. The latter sensitivity range is not yet limited by the uncertainty of a frequency standard which can provide a relative stability of ≈ $10^{-14}$ that minimizes possible sampling rate jitter and drifts ( $\delta B_{clock} < 10^{-5} \, fT$ ). Since the CRLB power law is based on statistical signal processing theory, noise sources inherent to the magnetometer may limit the ultimate sensitivity. To check this, the Allan Standard Deviation of the relative Larmor frequencies or phases of co-located $^3$He and $^{129}$Xe sample spins were analyzed showing the expected ~ $1/\sqrt{\tau}$ behavior for a observation time of T≈31000 s. In clock comparison experiments of this type, the sensitivity to magnetic field fluctuations cancels to the first order, a situation which is met in atomic clocks, too, i.e., in transitions between two specific hyperfine levels of the ground state (Cs-atomic clock, for example). Therefore, the detection of the free spin precession of co-located $^3$He/$^{129}$Xe spin samples can be used as ultra-sensitive probe for non-magnetic spin interactions , like in searches for a Lorentz-violating sidereal modulation of the precession frequency or in searches for spin-dependent short-range interactions induced by light, pseudoscalar bosons such
as the axion invented to solve the strong CP problem [54].

In our paper we presented two actual applications of low-field magnetometry based on the detection of free spin precession: The first shows the use of a $^3$He magnetometer in novel neutron EDM experiments in years to come, where the precise knowledge (i.e., better than 10 fT) of the average magnetic flux over large areas > 700 cm$^2$ is demanded. The sensitivity obtained with a prototype of a flat cylindrical magnetometer vessel reaches ≈ 2fT during one Ramsey cycle (T = 200s). The other application is a $^3$He/$^{129}$Xe clock-comparison experiment in which a high intrinsic frequency stability during a period of day is a prerequisite in order to detect tiny violations of Lorentz invariance as the laboratory reference frame (Earth) rotates with respect to a hypothetical background field, e.g., a potential field fixed to the rest frame of the cosmic microwave background. It is shown that a co-located $^3$He/$^{129}$Xe atomic clock can reach a measurement sensitivity of < 1 nHz.

There is still room for improvements: At present, the relatively short $T_{1,wall}^{Xe}$ relaxation time of $^{129}$Xe measured to be $T_{1,wall}^{Xe} \approx 5h$ limits the total observation time $T$ in our $^3$He/$^{129}$Xe clock comparison experiments based on free spin precession. Efforts to increase $T_{1,wall}^{Xe}$ considerably are therefore essential.

## ACKNOWLEDGMENTS

We would like to acknowledge W. Hoffmann, H. Pfeiffer, W. Riedel and R. Seemann from PTB for engineering electronic device. This work was supported by the Deutsche Forschungsgemeinschaft (DFG) under contract number BA 3605/1-1.
We are grateful to our glass blower R.Jera for preparing the low-relaxation glass vessels from GE180.

## Appendix A: A lower bound on the sensitivity to the frequency of an exponentially damped sinusoidal signal.

In the following we examine the determination of the Cramer-Rao Lower Bound (CRLB) for the frequency $f$ of an exponentially damped sinusoidal signal with amplitude $A$, damping factor $\beta$, and phase $\Phi$ embedded in a white Gaussian noise $w[n]$. The recorded data are assumed to be

$$s[n] = A \cdot \cos(2\pi \cdot f \cdot \Delta t \cdot n + \Phi) \cdot \exp(-\beta \cdot n) + w[n] \qquad n = 0,1,2,3,..., N-1 \qquad (A.1)$$

with $\Delta t = 1/r_s = 1/(2 f_{BW})$ and $\beta = \Delta t / T_2^*$.

We follow the notation in ref. [7]: The elements of the Fisher information matrix $[I(\Theta)]_{ij}$ are given by

$$[I(\Theta)]_{ij} = \frac{1}{\sigma^2} \sum_{n=0}^{N-1} \frac{\partial s[n;\Theta]}{\partial \Theta_i} \frac{\partial s[n;\Theta]}{\partial \Theta_j} \qquad (A.2)$$

where $\sigma^2$ is the variance of the white Gaussian noise and $\Theta = (A, f, \Phi, \beta)$ are the estimators of the different variables. Let $\alpha_n = 2\pi \cdot f \cdot \Delta t \cdot n + \Phi$ be the phase of the n$^{th}$ data point. One can show that we have for matrix elements

$$\propto \sum_{n=0}^{N-1} n^m \cos 2\alpha_n \cdot \exp(-2 \cdot \beta \cdot n) \approx 0 \text{ and } \propto \sum_{n=0}^{N-1} n^m \sin 2\alpha_n \cdot \exp(-2 \cdot \beta \cdot n) \approx 0 \qquad (A.3)$$

with m=0,1,2.

The Fisher information matrix then reads:

$$I(\Theta) = \begin{pmatrix} I_{11} & 0 & 0 & I_{14} \\ 0 & I_{22} & I_{23} & 0 \\ 0 & I_{32} & I_{33} & 0 \\ I_{41} & 0 & 0 & I_{44} \end{pmatrix} \qquad (A.4)$$

The matrix elements relevant to determine the frequency estimator $\sigma_f^2$ are:

$$I_{22} = \frac{1}{2\sigma^2}(2\pi A \cdot \Delta t)^2 \sum_{n=0}^{N-1} n^2 \exp(-2\beta n) \quad (A.5)$$

$$I_{33} = \frac{1}{2\sigma^2} A^2 \sum_{n=0}^{N-1} \exp(-2\beta n) \quad (A.6)$$

$$I_{23} = I_{32} = \frac{1}{2\sigma^2} 2\pi \cdot A^2 \cdot \Delta t \sum_{n=0}^{N-1} n \exp(-2\beta n) \quad (A.7)$$

Upon inversion $I(\Theta) \circ I^{-1}(\Theta) = 1$, we get the estimators from the diagonal elements of the Fisher inversion matrix $I^{-1}(\Theta)$. The frequency estimator $\sigma_f^2$ is then given by

$$\sigma_f^2 \geq I_{22}^{-1} = \frac{I_{33}}{I_{22} \cdot I_{33} - I_{23} \cdot I_{32}} \quad (A.8)$$

The sums over $n$ in equations (A5) and (A7) can be carried out by using the identities

$$\sum_{n=0}^{N-1} n \exp(-2\beta n) = -\frac{d}{d\beta}\left(\frac{1}{2}\sum_{n=0}^{N-1} \exp(-2\beta n)\right) \quad (A.9)$$

and

$$\sum_{n=0}^{N-1} n^2 \exp(-2\beta n) = \frac{d^2}{d\beta^2}\left(\frac{1}{4}\sum_{n=0}^{N-1} \exp(-2\beta n)\right) \quad (A.10)$$

Putting in the expressions from equation (1), replacing $\sigma^2$ by $N_\alpha^2$, and using $T = \Delta t \cdot N$, we finally get

$$\sigma_f^2 \geq \frac{12}{(2\pi)^2 (A/\rho_\alpha)^2 \cdot T^3} \cdot C \quad (A.11)$$

For C=1, we reproduce the frequency estimation for a pure sinusoidal signal (see Eq.(3)). The effect of exponential damping enters in the second factor (C >1):

$$C = \frac{N^3}{12} \cdot \frac{(1-z^2)^3 \cdot (1-z^{2N})}{z^2 \cdot (1-z^{2N})^2 - N^2 z^{2N}(1-z^2)^2} \quad (A.12)$$

with $z = \exp(-\beta)$.

Figure A1 shows the effect of the exponential damping ($\sqrt{C}$) for observation times $0 \leq T \leq 10000s$ assuming a $T_2^*$ of $T_2^* = 10000s$.

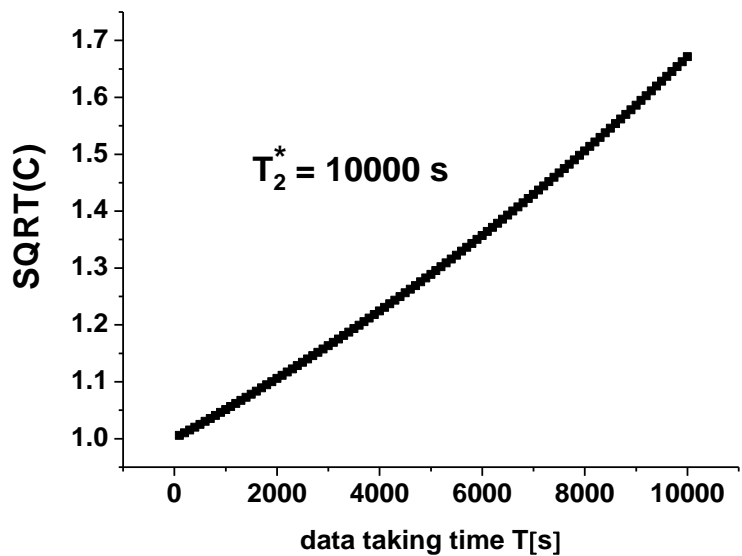

Fig.A1: Deviation from the $\sqrt{\sigma_f^2} \propto 1/T^{3/2}$ CRLB power law for a sinusoidal signal in case of an exponentially damped signal. $T_2^*$ is assumed to be $T_2^* = 10000s$.

## References:

[1] SQUID Sensors: Fundamentals, Fabrication and Applications, edited by H.Weinstock, (Kluwer Academic,Dordrecht, 1996)

[2] E.B. Aleksandrov, M.V. Balabas, A.K. Vershovskii, and A.S. Pazgalev, Tech. Phys.**49,** 779 (2004)

[3] D. Budker, D.F. Kimball, S.M. Rochester, V.V. Yashchuk, and M. Zolotorev, Phys. Rev. A **62,** 043403 (2000)

[4] S. Groeger, G. Bison, J.L. Schenker, R. Wynands, and A. Weis, Eur. Phys. J. D **38,** 239 (2006)

[5] I.K. Kominis, T.W. Kornack, J.C. Allred, and M.V. Romalis, Nature **422,** 596 (2003)

[6] C. Cohen-Tannoudji, J. DuPont-Roc, S. Haroche, and F. Laloë, Phys. Rev.Lett. **22,**758 (1969)

[7] S.M. Kay, *Fundamentals of Statistical Signal Processing: Estimation Theory,* Vol I, (Prentice Hall, New Jersey, 1993)

[8] J. Allred, R. Lyman, T. Kornack, and M.V. Romalis, Phys. Rev. Lett. **89,** 130801 (2002)

[9] J.A.Barnes et al., IEEE Trans. Instrum. Meas. **20**,105 (1971)

[10] G. Tastevin, S. Grot, E. Courtade, S. Bordais and P.-J. Nacher, Appl.Phys.B **78**,145 (2004)

[11] H. Zhu, I. C. Ruset, and F. W. Hersman, Opt. Lett. **30**, 1342 (2005)

[12] M.Wolf, Ph.D. thesis, University of Mainz, 2004

[13] D. Drung, Physica C **368**,134 (2002)

[14] W. Kilian, A. Haller, F. Seifert, D. Grosenick, and H. Rinneberg, Eur. Phys. J. D **42**, 197 (2007)

[15] M. Burghoff, S. Hartwig, W. Kilian, A. Vorwerk and L. Trahms, IEEE Trans. App. Supercon. **17**, 846 (2007)

[16] C.P.Slichter, *Principles of Magnetic Resonance*, 3$^{rd}$ edn. (Springer, Berlin, 1996)

[17] G.D. Cates, S.R. Schaefer, and W. Happer, Phys. Rev. A **37**,2877 (1988)

[18] D.D. McGregor, Phys. Rev. A **41**, 2631(1990)

[19] R. Barbé, M. Leduc, and F. Laloë, J.Phys.France **35**, 935 (1974);

Landolt-Börnstein, *Diffusion in Gases, Liquids* and *Electrolytes*, Part A (Springer, Berlin, 2007)

[20] S.N. Erné, H.D. Hahlbohm, H. Scheer, Z. Trontelj, *The Berlin Magnetically Shielded Room - Performances*, in *Biomagnetism*, edited by S.N. Erné, H.D. Hahlbohm, H. Lübbig (Walter de Gruyter, Berlin, New York, 1981), p.79

[21]J. Bork, H.-D. Hahlbohm, R. Klein, and A. Schnabel, Proc. Biomag **2000**, 970 (2000)

[22] F.Thiel, A.Schnabel, S.Knappe-Grüneberg, D. Stollfuß, and M. Burghoff, Rev. Sci. Instr. **78** , 035106 (2007)

[23] D.Drung, Supercond. Sci. Technol. **16** , 1320 (2003)

[24] L.D. Schearer, F.D. Colegrove, and G.K. Walters, Phys. Rev. Lett. **10**, 108 (1963)

[25] A. Schnabel, M. Burghoff, S. Hartwig, F. Petsche, U. Steinhoff, D. Drung, H. Koch, Neurology and Clinical Neurophysiology, 2004:70 (November 30, 2004)

[26] M. Burghoff, A. Schnabel, D. Drung, F. Thiel, S. Knappe-Grüneberg, S. Hartwig, O. Kosch, L. Trahms, H. Koch, Neurology and Clinical Neurophysiology, 2004:67 (November 30, 2004)

[27] W. Kilian, Ph.D. thesis, Freie Universität Berlin, 2001; www.diss.fu-berlin.de

[28] J. Schmiedeskamp, W. Heil, E.W. Otten, R.K. Kremer, A. Simon, and J. Zimmer, Eur. Phys. J. D **38** , 427 (2006)

[29] A. Deninger, W. Heil, E.W. Otten, M. Wolf, R.K. Kremer, and A. Simon, Eur. Phys. J. D **38** , 439 (2006)

[30] J. Schmiedeskamp, H.-J. Elmers, W. Heil, E.W. Otten, Yu. Sobolev, W. Kilian, H. Rinneberg,T. Sander-Thömmes, F. Seifert, and J. Zimmer, Eur. Phys. J. D **38** , 445 (2006)

[31] N.P. Bigelow, P.J. Nacher, and M. Leduc, J. Physique II **2** , 2159 (1992) 2159

[32] Neutron EDM Collaboration at PSI ; http://nedm.web.psi.ch/index.htm

[33] K. Green, P.G. Harris, P. Iaydjiev , et al., Nucl. Instr. and Meth. A **404**, 381 (1998)

[34] M. Pendlebury , W. Heil, Yu. Sobolev et al., Phys. Rev. A **70**, 032102 (2004)

[35] S. Lamoreaux and R. Golub, Phys. Rev. A **71**, 032104 (2005)

[36] N.F. Ramsey, Acta.Phys.Hungar.**55**, 117 (1984)

[37] N.F.Ramsey, *Molecular Beams*, Oxford University Press, (1956)

[38] V.W. Hughes, H.G. Robinson, and V. Beltran-Lopez, Phys. Rev. Lett. **4** ,342 (1960)

[39] J.D. Prestage, J.J. Bollinger, W.M. Itano, and D.J. Wineland, Phys. Rev. Lett. **54** , 2387 (1985)

[40] S.K. Lamoreaux, J.P. Jacobs, B.R. Heckel, F.J. Raab, and E.N. Fortson, Phys. Rev. Lett. **57**, 3125 (1986)

[41] T.E. Chupp, R.J. Hoare, R.A. Loveman, E.R. Oteiza, J.M. Richardson, and M.E. Wagshul, Phys. Rev. Lett. **63**, 1541 (1989)

[42] C.J. Berglund, L.R. Hunter, D. Krause, E.O. Prigge, and M.S. Ronfeldt, Phys. Rev. Lett. **75** , 1879 (1995)

[43] M.A. Rosenberry and T.E. Chupp, Phys. Rev. Lett. **86** , 22 (2001)

[44] M.V. Romalis, W.C. Griffith, J.P. Jacobs, and E.N. Fortson, Phys. Rev. Lett. **86** , 2505 (2001)

[45] L.R. Hunter et al., in *CPT and Lorentz Symmetry*, edited by V.A. Kostelecky (World Scientific, Singapore,1999)

[46] V.A. Kostelecky and Ch.D. Lane, Phys. Rev. D **60**, 116010 (1999)

[47] D. Bear, Ch.D. Lane, V.A. Kostelecky, R.E. Stoner, and R.L.Walsworth, Phys. Rev.Lett. **85** ,5038 (2000)

[48] B. Chann, I.A. Nelson, L.W. Anderson, B. Driehuys, and T.G. Walker, Phys.Rev.Lett. **88,** 113201 (2002)

[49] R.W. Mair, P.N. Sen, M.D. Hurlimann, S. Patz, D.G. Cory, and R.L. Walsworth, J.Magn.Res. **156**, 202 (2002) and references therein

[50] R.H. Acosta, L. Agulles-Pedrós, S. Komin, D. Sebastiani, H.W. Spiess, and P. Blümler, Physical Chemistry Chemical Physics **8**, 4182 (2006)

[51] K.C. Hasson, G.D. Cates, K. Lerman, P. Bogorad, and W. Happer, Phys. Rev.A **41,** 3672 (1990)

[52] International Council for Science: Committee on Data for Science and Technology (CODATA), www.codata.org , 2007

[53] M. Pfeiffer and O. Lutz, J. Magn. Res. A **108** , 106 (1994)

[54] J.E. Moody and Frank Wilczek, Phys. Rev. D **30** (1984) 130.